\documentclass[journal=jctcce,manuscript=article]{achemso}
\usepackage{amsmath}
\usepackage{amssymb}
\usepackage{amsfonts}
\usepackage{units}
\usepackage{bm}
\usepackage{float}
\usepackage[utf8]{inputenc}
\usepackage[colorlinks=true,citecolor=blue,linkcolor=blue]{hyperref}
\usepackage{pdfpages}

\renewcommand{\vec}[1]{\ensuremath{\boldsymbol{#1}}}
\newcommand{\rdep}{\left(\vec{r}\right)}
\DeclareMathOperator\erf{erf}

\author{Tom\'{a}\v{s} Rauch} 
\affiliation{Institut für Festkörpertheorie und -optik, Friedrich-Schiller-Universität Jena, Max-Wien-Platz 1, 07743 Jena, Germany}
\email{tomas.rauch@uni-jena.de}

\author{Miguel A. L. Marques} 
\affiliation{Institut für Physik, Martin-Luther-Universität Halle-Wittenberg, 06120 Halle/Saale, Germany}
\alsoaffiliation{European Theoretical Spectroscopy Facility}

\author{Silvana Botti} 
\affiliation{Institut für Festkörpertheorie und -optik, Friedrich-Schiller-Universität Jena, Max-Wien-Platz 1, 07743 Jena, Germany}
\alsoaffiliation{European Theoretical Spectroscopy Facility}

\title{Local modified Becke-Johnson exchange-correlation potential for interfaces, surfaces, and two-dimensional materials}

\begin{document}

%
%

\begin{abstract}
 The modified Becke-Johnson meta-GGA potential of density functional theory has been shown to be the best exchange-correlation potential to determine band gaps of crystalline solids. However, it cannot be consistently used for the electronic structure of non-periodic or nanostructured systems. We propose an extension of this potential that enables its use to study heterogeneous, finite and low-dimensional systems. This is achieved by using a coordinate-dependent expression for the parameter $c$ that weights the Becke-Russel exchange, in contrast to the original global formulation, where $c$ is just a fitted number. Our potential takes advantage of the excellent description of band gaps provided by the modified Becke-Johnson potential and preserves its modest computational effort. Furthermore, it yields with one single calculation band diagrams and band offsets of heterostructures and surfaces. We exemplify the usefulness and efficiency of our local meta-GGA potential by testing it for a series of interfaces (Si/SiO$_2$, AlAs/GaAs, AlP/GaP, and GaP/Si), a Si surface, and boron nitride monolayer.
\end{abstract}

\section{Introduction}
The world around us is inhomogeneous. Approximately homogeneous parts of matter are separated from each other and from the surrounding vacuum by interfaces and surfaces, and these regions are at the origin of a vast number of fascinating and useful phenomena studied across different fields, ranging from biology, over soft matter, to solid-state physics~\cite{Allara2005}. The photoelectric effect, quantum Hall effect, symmetry-protected topological states, or the electron flow in a transistor are well known examples of physics emerging at interfaces or surfaces~\cite{Brillson2012}. Understanding electronic properties such as local band gaps, band alignments, energy levels of localized states at interfaces and surfaces is crucial to interpret and control phenomena arising in these regions, opening the way to technological breakthroughs.

The most successful method for the theoretical study of surfaces and interfaces in unquestionably density-functional theory (DFT)~\cite{Hohenberg1964,Kohn1965}. This theory combines an unparalleled accuracy with relatively mild computational requirements. In Kohn-Sham DFT all complexities of the many-electron system are included in the so-called exchange-correlation (XC) functional. This is a rather complicated quantity that has to be approximated in any practical use of DFT, and that ultimately determines the quality of the results. Standard semilocal approximations to the XC functionals are quite successful in predicting many properties of solids, such as the atomic structure, phonon spectra or the qualitative band structure. Unfortunately, for an accurate description of band gaps and band alignments, it is necessary to use more advanced approximations, like hybrid functionals \cite{Seidl1996,Scuseria2005} or even many-body $GW$ methods~\cite{Hedin1965,Onida2002}. These are computationally much more expensive and can be applied to small surface or interface models only. However, these small models are often not good enough to approximate the inhomogeneous regions of real systems.

A possible way out of this vicious circle is the usage of meta-GGA functionals. Here we will be interested in one meta-GGA functional, the modified Becke-Johnson (MBJ) XC potential~\cite{Tran2009}, which has been optimized for the description of electronic band gaps of homogeneous solids. Various comparisons~\cite{Singh2010,Koller2011,Tran2017,Borlido2019} show that the MBJ is the best semilocal approximation to determine band gaps, achieving on average an accuracy even better than the one of hybrid functionals~\cite{Borlido2019}, and at a much lower computational price.

\section{Original modified Becke-Johnson potential}
The modified Becke-Johnson exchange-correlation potential was proposed by Tran and Blaha in 2009~\cite{Tran2009}. It belongs to the meta-GGA family and its exchange part has the form
\begin{equation}
    \label{eq:MBJ_def}
    v^{\mathrm{MBJ}}_{x}\rdep = c v^{\mathrm{BR}}_{x}\rdep + \left(3 c - 2\right) \frac{1}{\pi} \sqrt{\frac{5}{12}}\sqrt{\frac{2 t \rdep}{\rho \rdep}} \,,
\end{equation}
where $\rho\rdep = \sum^{N}_{i} \left|\psi_{i}\rdep\right|^2$ is the electronic density, $t\rdep = \sum^{N}_{i} \nabla\psi^{*}_{i} \cdot \nabla\psi_{i}$ the kinetic-energy density and
\begin{equation}
    \label{eq:BR_def}
    v^{\mathrm{BR}}_{x}\rdep = -\frac{1}{b\rdep}\left( 1-e^{-x\rdep} - \frac{1}{2}x\rdep e^{-x\rdep} \right)
\end{equation}
is the Becke-Russel (BR) exchange potential~\cite{Becke1989}, with $x\rdep$ being calculated from $\rho\rdep$ and its spatial gradient and Laplacian and \mbox{$b\rdep = \sqrt[3]{x^3 e^{-x}/(8\pi\rho\rdep)}$}. Together with the LDA correlation, as proposed originally, we will refer to the whole potential as the MBJ XC potential.

The original Becke-Johnson (BJ) potential~\cite{Becke2006} (obtained with $c=1$ in Eq.~\ref{eq:MBJ_def}) was proposed as a sum of the Slater potential~\cite{Slater1951} describing a Coulomb potential of an exchange hole and $\frac{1}{\pi} \sqrt{\frac{5}{12}}\sqrt{\frac{2 t \rdep}{\rho \rdep}}$, which corrects the error of the Slater potential with respect to the exactly solvable ``optimized effective potential''~\cite{Sharp1953} for atoms. 
Becke and Johnson have further shown that using the BR potential instead of the Slater potential gives almost identical results for atoms~\cite{Becke2006}. In the MBJ potential (Eq.~\ref{eq:MBJ_def}) Tran and Blaha used the BR potential, and they introduced a ``mixing parameter'' $c$ reminiscent of the mixing in hybrid functionals. Indeed, the BJ potential models ``exact exchange'' and the term proportional to $t \rdep/\rho \rdep$ can be seen as a screening, thus justifying the analogy to hybrid functionals even further. Led by this analogy and the idea of material dependent mixing~\cite{Krukau2008}, Tran and Blaha proposed to determine $c$ as 
\begin{equation}
    \label{eq:c_def}
    c = \alpha + \beta \bar{g}^\epsilon
\end{equation}
with
\begin{equation}
    \label{eq:g_def}
    \bar{g} = \frac{1}{V_{\mathrm{cell}}} \int_{\mathrm{cell}} d^{3}r\ \frac{\left|\nabla \rho\rdep\right|}{\rho\rdep},
\end{equation}
averaging $g\rdep=\left|\nabla\rho\rdep\right|/\rho\rdep$ in a unit cell.
Originally~\cite{Tran2009}, the exponent $\epsilon$ in Eq.~\eqref{eq:c_def} was fixed to $1/2$ and the parameters $\alpha$ and $\beta$ were fitted to a set of materials, to minimize the error with respect to their experimental band gaps. Later, an improved fit was performed for $\epsilon=1$ resulting in $\alpha=0.488$ and $\beta=\unit[0.5]{bohr}$~\cite{Koller2012}. We chose the latter parameters for this work, as they predict band gaps of semiconductors very close to their experimental values. Other parameters were also obtained recently for more specialized material sets~\cite{Jishi2014,Traore2019}.

In spite of its many virtues, analyzed in detail in Ref.~\citenum{Koller2011}, the MBJ potential also suffers from drawbacks. For example, it is not a functional derivative of any density functional~\cite{Karolewski2009,Gaiduk2009}, and therefore it  violates a few exact conditions~\cite{Levy1985} and it cannot be used to calculate total energies. Yet another, more practical, problem originates from the form of Eq.~\eqref{eq:g_def}. Since $g$ is averaged over the whole periodic unit cell, the potential can not be consistently used for inhomogeneous systems. This is better explained with a couple of examples. Let us consider a heterostructure made of two materials with very different values of $c$. In this case, the MBJ potential would use a value of $c$ averaged over the whole supercell, leading to an incorrect description of the local band gaps of both constituents. Another example are low-dimensional systems, such as surfaces or molecules. In these cases, $\bar{g}$ converges with the size of the unit cell to a completely inadequate value that depends on the ionization potential of the system. Some groups tried to solve this problem either by fixing the value of $c$ to the one of the bulk~\cite{Li2015}, or by constraining the size of the vacuum region to the value that yields the bulk $c$ parameter~\cite{Smith2014}.
These procedures might in some cases result in good band gaps in bulk-like regions, but the quality of the description of surfaces is highly questionable. 

The impossibility to describe reliably the electronic structure of heterostructures or finite systems is a serious drawback that hampers the systematic application of this meta-GGA potential to evaluate band gaps or band diagrams in high-throughput calculations for computational materials design. For such calculations, the state-of-the-art for band structures remains the more expensive screened hybrid functional HSE06~\cite{HSE03,HSE06}, despite its significantly higher computational cost and larger mean average error~\cite{Borlido2019}. Here, we propose an effective solution to enable the use of the MBJ potential in automated calculations of nanostructured systems, through an inexpensive local reformulation of the parameter $c$.

\section{Formulation of the local MBJ potential} 

We extend the scheme that we had originally applied to obtain a local hybrid functional for interfaces~\cite{Marques2011,Borlido2018} and we define the locally averaged but spatially varying function
\begin{equation}
    \label{eq:g_loc}
    \bar{g}\rdep = \frac{1}{\left(2\pi\sigma^2\right)^{3/2}} \int d^3 r'\ g\left(\vec{r}'\right)\  e^{-\frac{\left|\vec{r}-\vec{r}'\right|^2}{2\sigma^2}}
\end{equation}
that depends on a smearing parameter $\sigma$. We will discuss in the following how $\sigma$ can be determined once and for all, and set as a parameter that defines the potential. The possibility to use a smeared local estimator was suggested by Marques {\it et al.}~\cite{Marques2011} and mentioned as a promising perspective in Refs.~\cite{Koller2012,Tran2015}, but no realization had been attempted yet. The form of $\bar{g}$ in Eq.~\eqref{eq:g_loc} is particularly convenient because it can be easily implemented into DFT codes using fast-Fourier transforms via a convolution of $g\rdep$ and the Gaussian in the reciprocal space. We thus introduce the local MBJ (LMBJ) exchange potential with the local parameter $c\rdep$ given by
\begin{equation}
    \label{eq:c_loc}
    c\rdep = \alpha + \beta \bar{g}\rdep
\end{equation}
with $\bar{g}\rdep$ as in Eq.~\eqref{eq:g_loc} and $\alpha=0.488$ and $\beta=\unit[0.5]{bohr}$.

In principle, the LMBJ potential with the local estimator (Eq.~\ref{eq:g_loc}) could be already used for surfaces and other systems with vacuum. However, a few problems remain. To recover the correct asymptotic behavior of the XC potential, it is necessary that $c\rightarrow 1$ in the vacuum region. Furthermore, at the matter-vacuum boundary $g\rdep=\left|\nabla\rho\rdep\right|/\rho\rdep$ takes values ranging up to $\sim $\unit[1000]{bohr$^{-1}$}, leading to extremely large values of the XC potential and thus hindering the calculation from converging. Another complication arises from the fact that $\rho\rdep$ becomes vanishingly small far from the nuclei, leading to numerical instabilities. We solve all these problems by enforcing $c\rdep \rightarrow 1$ for regions of low density through the modification
\begin{equation}
    \label{eq:g_vac}
    g\rdep = \frac{1-\alpha}{\beta}\left[1 - \erf\left(\frac{\rho\rdep}{\rho_{\mathrm{th}}}\right) \right]
    + \frac{\left|\nabla \rho\rdep\right|}{\rho\rdep} \erf\left(\frac{\rho\rdep}{\rho_{\mathrm{th}}}\right) 
\end{equation}
and by introducing a threshold density $\rho_{\mathrm{th}}$. For $\rho\rdep\gg \rho_{\mathrm{th}}$ we obtain the previous limit $g\rdep=\left|\nabla\rho\rdep\right|/\rho\rdep$, while in the opposite case $g\rdep=(1-\alpha)/\beta$ and $c\rdep =1$.

Equations~\ref{eq:g_loc} and \ref{eq:g_vac} define our LMBJ potential and have been implemented in the VASP code~\cite{Kresse1996}. Since the projector-augmented-waves (PAW) method is used~\cite{Kresse1999}, the implementation includes a careful treatment of the PAW spheres in addition to the plane-wave part. We give technical details on the implementation in the Supplemental Material.

Before applying the LMBJ potential to realistic systems, we have to choose appropriate values for the parameters $\sigma$ and $\rho_{\mathrm{th}}$. The length $\sigma$ should be as small as possible, to allow for an accurate description of local electronic properties, but also large enough to keep the properties of the original MBJ potential in locally homogeneous regions. We set $\sigma$ = \unit[3.78]{bohr} = \unit[2]{\AA},
which means that $g\rdep$ is averaged over a region which covers typical interatomic distances. We remark that a similar value was selected for the corresponding $\sigma$ parameter in Ref.~\citenum{Borlido2018}. As shown in Tab.~\ref{tab:bulk_gaps} and Fig. 1 of the Supplemental Material, with $\sigma$ = \unit[3.78]{bohr} the calculated band gap reaches its saturation value for all test materials, which in a turn recovers results obtained with the original MBJ potential ($\alpha=$ \unit[-0.012], $\beta=$ \unit[1.023]{bohr$^{1/2}$}, $\epsilon=$ 0.5).
%
\begin{table}[h!]
    \centering
    \begin{tabular}{|c|c|c|c|c|c|c|}
        \hline
         & $\sigma=0.95$, & $\sigma=3.78$, & $\sigma=5.67$, & $\sigma=3.78$, & &  \\
        material & $r^{\mathrm{th}}_s=5.0$ & $r^{\mathrm{th}}_s=5.0$ & $r^{\mathrm{th}}_s=5.0$ & $r^{\mathrm{th}}_s=2.0$ & MBJ & exp\\
        \hline
        Si & 1.11 & 1.20 & 1.20 & 1.18 & 1.27 & 1.17$^{a}$ \\
        $\beta$-SiO$_2$ & 8.85 & 8.63 & 8.59 & 7.88 & 8.13 & 10.30$^{b}$ \\
        AlAs & 1.93 & 2.12 & 2.13 & 2.10 & 2.15 & 2.23$^{a}$ \\
        GaAs & 1.59 & 1.61 & 1.61 & 1.59 & 1.59 & 1.52$^{a}$ \\
        AlP & 2.14 & 2.34 & 2.34 & 2.32 & 2.37 & 2.51$^{a}$ \\
        GaP & 2.39 & 2.38 & 2.39 & 2.35 & 2.38 & 2.35$^{a}$ \\
        \hline
        MAE & 0.38 & 0.35 & 0.36 & 0.47 & 0.43 &  \\
        \hline
        MAPE & 0.09 & 0.06 & 0.06 & 0.07 & 0.07 &  \\
        \hline
    \end{tabular}
    \caption{Band gap values (in eV) of bulk materials obtained with LMBJ potential for different values of $\sigma$ and $r^{\mathrm{th}}_s$ (both in bohr) compared with the original MBJ potential and experimental values given in $^{a}$~Ref.~\citenum{Steiner2014} and $^{b}$~Ref.~\citenum{Marques2011}. Mean average error (MAE) and mean average percentage error (MAPE) are also given. Note that due to the small test set, errors appear much smaller than from a large benchmark~\cite{Borlido2019}.}
    \label{tab:bulk_gaps}
\end{table}{}

Concerning the threshold density $\rho_{\mathrm{th}}$, we chose a value corresponding to the threshold Wigner-Seitz radius $r^{\mathrm{thr}}_s = (3/4\pi\rho_{\mathrm{th}})^{(1/3)} =$~\unit[5]{bohr}. This value lies well above the $r_s$ value of most of the metals listed in Ref.~\citenum{Ashcroft1976} and our tests show that we obtain bulk band gaps of common semiconductors very close to those yielded by the MBJ potential, see Tab.~\ref{tab:bulk_gaps}. Clearly, a different choice should be made if one is interested in simulating materials with extremely low electronic densities. Setting $r^{\mathrm{thr}}_s=$~\unit[2]{bohr} increases the deviation from the MBJ band gaps only slightly and leads to an overall underestimation of the band gaps.

Note that further optimization of $r^{\mathrm{thr}}_s$ is not trivial, as it requires a sufficiently large test set with reliable electronic properties for low-dimensional materials. Alternatively, the parameter $r^{\mathrm{thr}}_s$, possibly together with $\alpha$ and $\beta$, could be fitted to atomic or molecular properties, such as the ionization potential. Of course, the challenge in this case is to recover the predictive power of the MBJ potential for bulk semiconductors.

\section{Application of the LMBJ potential} 
We are going to apply now the LMBJ potential to study band diagrams of electronic systems in which the crystal periodicity is broken in one direction (e.g., $z$): we therefore consider the local value of $c\rdep$ averaged in the $xy$-plane, $\overline{c}_{xy}(z)$, and we calculate the local density of states (LDOS) along the z axis, see the Supplemental Material for details.

An important test of our potential is the Si/SiO$_2$ interface, since the bulk $c$ values and the bulk band gaps of the two constituents differ significantly. For this system, we expect that the standard MBJ potential with an averaged $c$ value leads to a poor description of the band gaps of both Si and SiO$_2$. To make a direct comparison with calculations in literature possible, we use the same supercell generated by Giustino and Pasquarello~\cite{Giustino2005}, already used in Refs.~\cite{Shaltaf2008,Borlido2018}. The interface model consists of 11 Si atomic layers along the (001) direction and 10 layers of SiO$_2$ in the $\beta$-cristobalite form. We adopted an $8\times 8\times 2$ $\mathbf{k}$-point grid and a cut-off energy of \unit[400]{eV}. As in all other calculations in this work, we used PAW pseudopotentials~\cite{Kresse1999} and a spin unpolarized formalism. We first preconverged the calculation using the Perdew-Burke-Ernzerhof (PBE) functional~\cite{Perdew1996} and used the result as a starting point for the subsequent LMBJ calculation. 
\begin{figure} 
  \centering
  \includegraphics[trim={5 0 25 0},clip,width = 0.99\columnwidth]{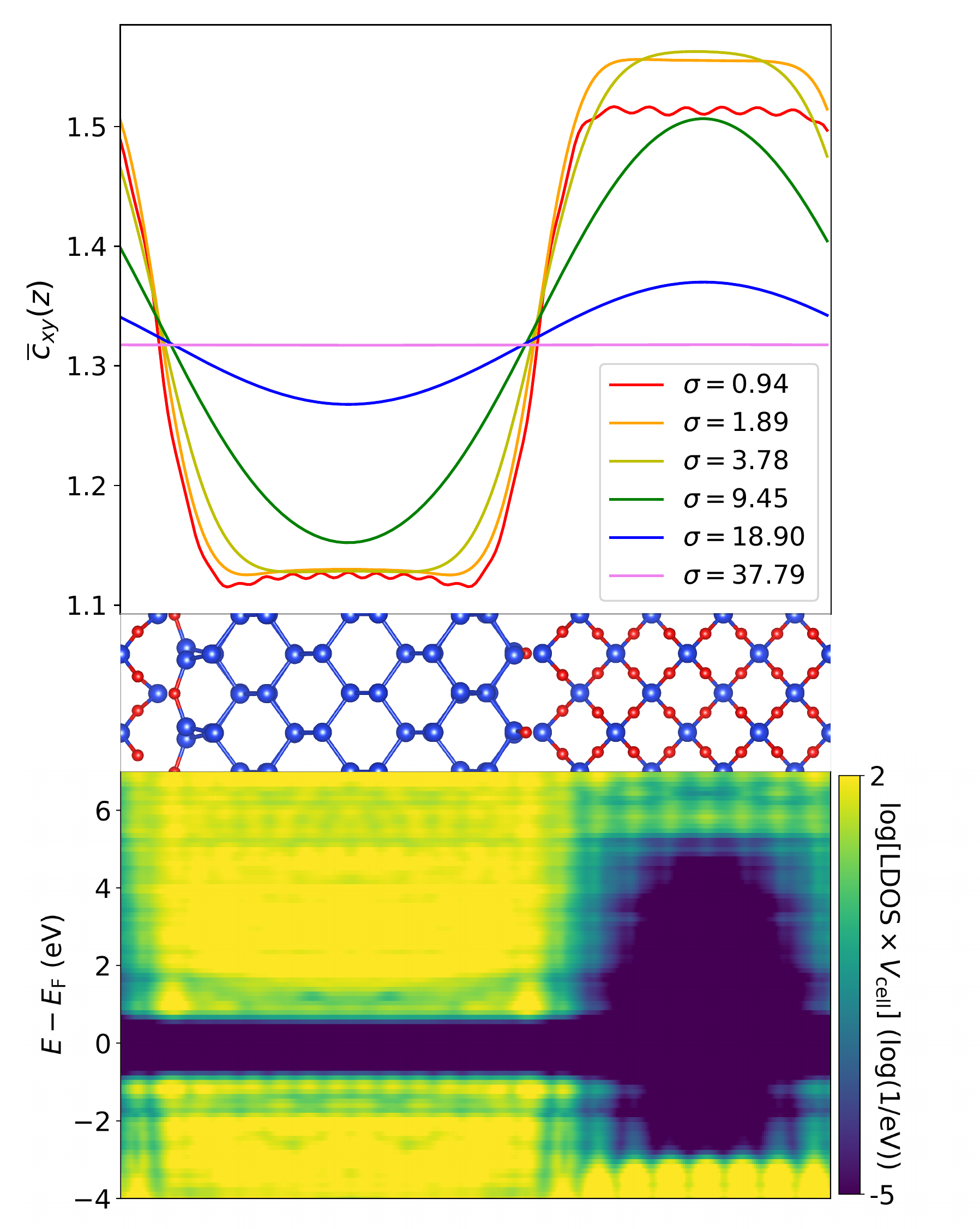}
  \caption{Band diagram of the Si/SiO$_2$ heterostructure. Top: Averaged $\overline{c}_{xy}(z)$ for different values of smearing $\sigma$ given in bohr. Middle: Atomic structure of the interface model (drawn with VESTA~\cite{Momma2011}), where blue depicts Si and red represents O atoms. Bottom: Logarithm of \mbox{$\mathrm{LDOS} \times V_{\mathrm{cell}}$} averaged in the $xy$-plane (yellow: high LDOS, violet: low LDOS) calculated with $\sigma$ = \unit[3.78]{bohr} on a $2\times 2\times 1$ $\mathbf{k}$-point grid.}
  \label{fig:fig_heterostructure}
\end{figure}
\begin{figure*} 
  \centering
  \includegraphics[width = 0.30\textwidth]{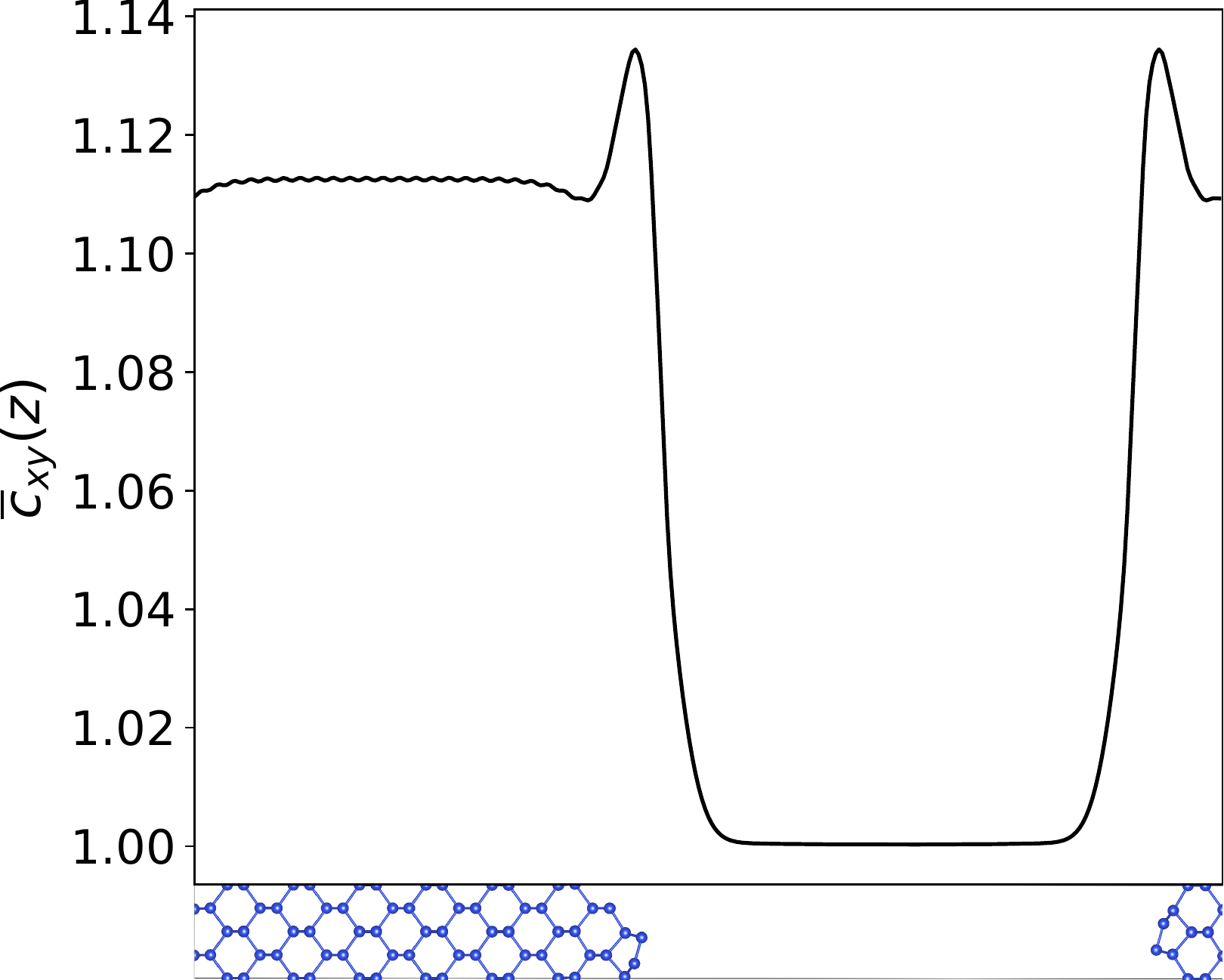}
  \includegraphics[width = 0.335\textwidth]{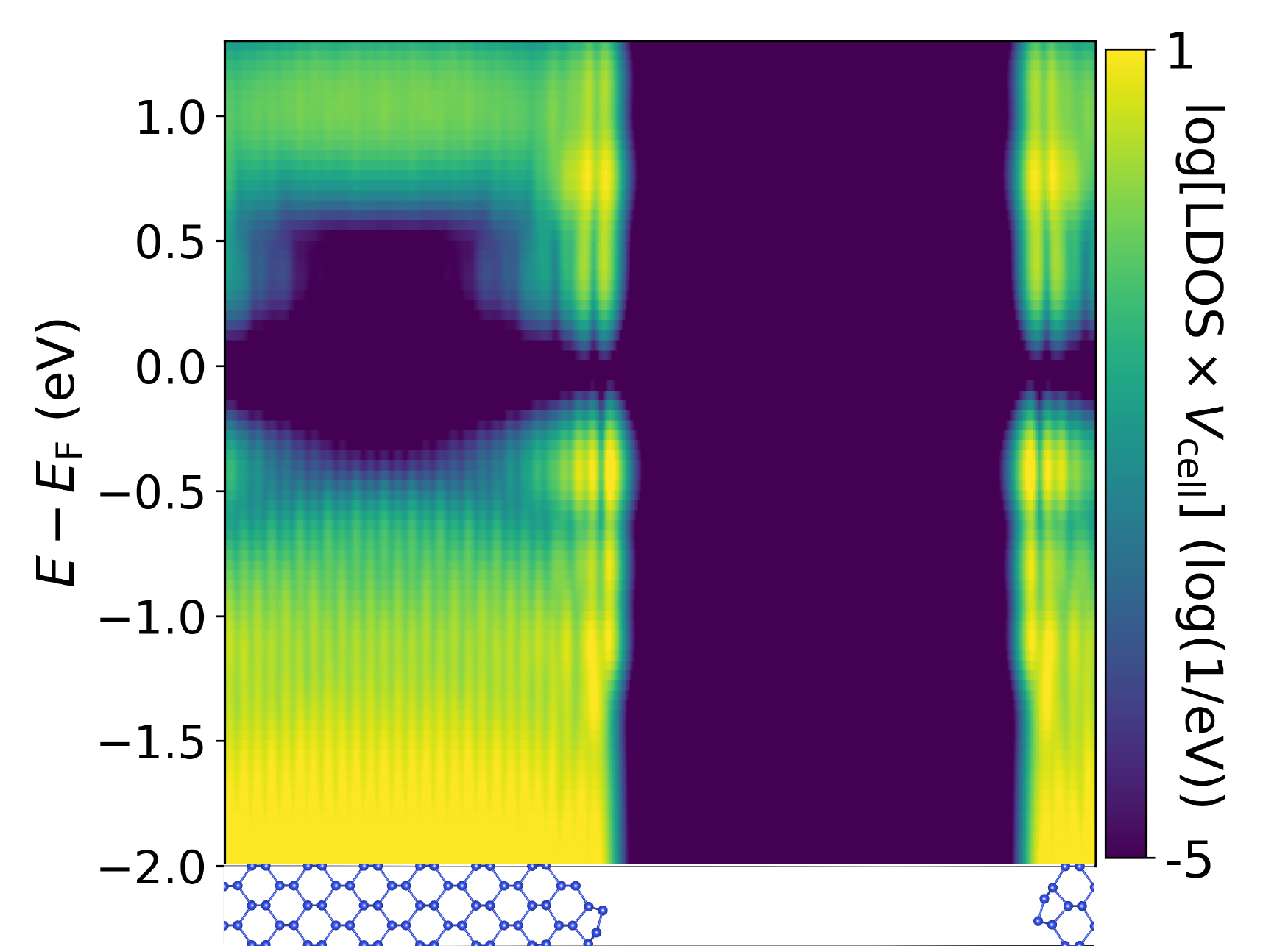}
  \includegraphics[width = 0.345\textwidth]{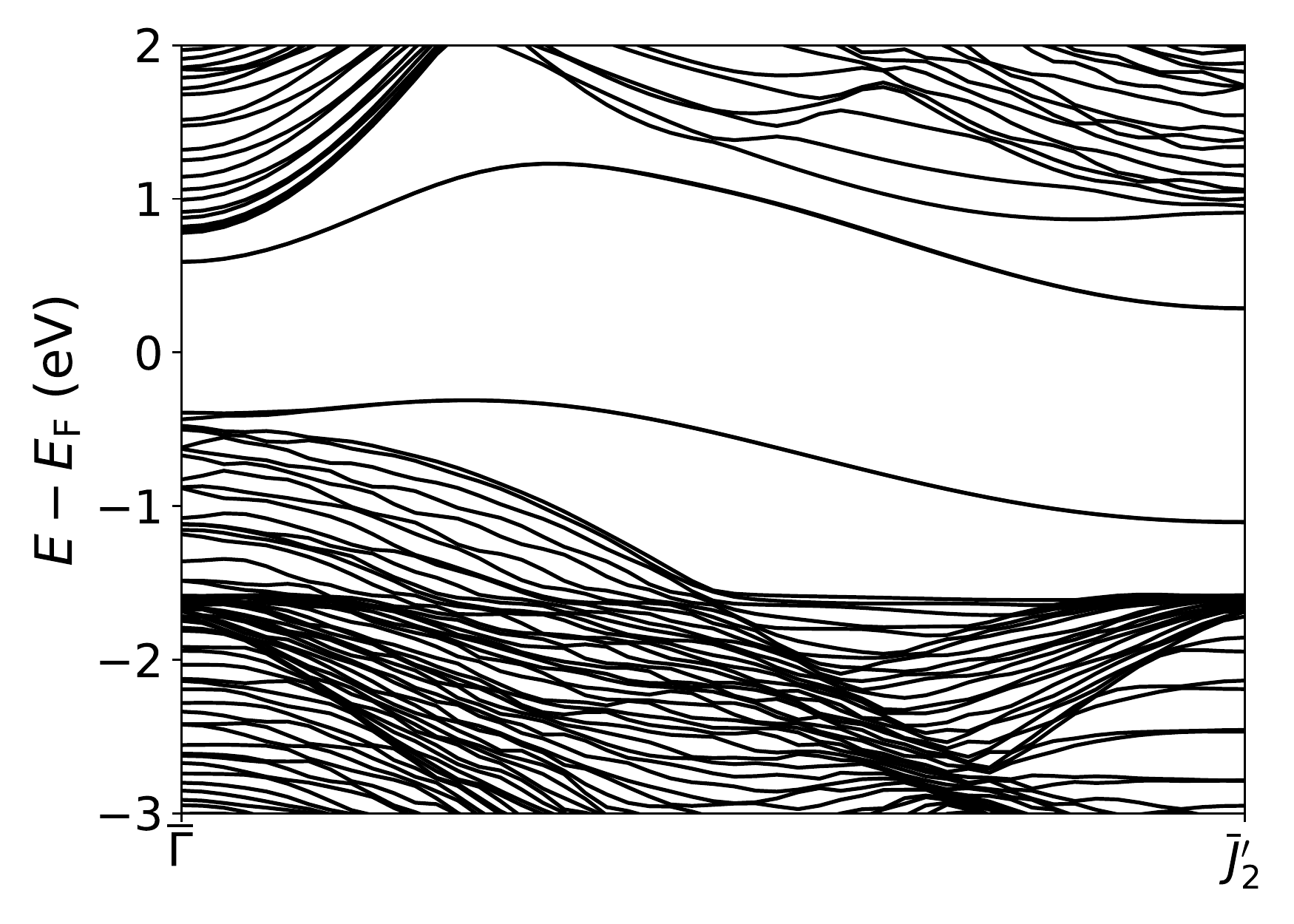}
  \caption{Clean Si(001) surface with $2\times 1$ reconstruction. Left: Averaged mixing $\overline{c}_{xy}(z)$. Middle: Logarithm of the LDOS averaged in the $xy$-plane. Right: Band structure of the slab along a line from the center $\overline{\Gamma} = (0,0)$ to the corner $\overline{J}'_{2} = (\pi/a,\pi/a)$ of the unreconstructed surface Brillouin zone.}
  \label{fig:fig_surf}
\end{figure*}

In the top panel of Fig.~\ref{fig:fig_heterostructure} we show the in-plane averaged local mixing for different values of the smearing $\sigma$. For the very large $\sigma =$~\unit[37.8]{bohr}, i.e., a smearing over the whole heterostructure, we obtain as expected a constant mixing $\overline{c}_{xy}(z)=1.32$. In this limit, our LMBJ potential restores the original MBJ (a reference calculation with the MBJ potential yields $c=1.30$). This calculation thus overestimates, underestimates the $c$-parameter for Si, SiO$_2$ respectively, and consequently the band gaps. Decreasing the smearing leads to a variation of $c$ across the slab. We find $\sigma$ = \unit[3.78]{bohr} to be optimal, since it reproduces well the bulk mixing of both Si (heterostructure: $c=1.13$, bulk: $c=1.11$) and SiO$_2$ (heterostructure: $c=1.56$, bulk: $c=1.58$). Interestingly, our optimal smearing value agrees with the one obtained in Ref.~\citenum{Borlido2018} for local hybrids using the same Gaussian smearing of the local mixing parameter applied to the same interface model. 

In the bottom panel of Fig.~\ref{fig:fig_heterostructure} we present the LDOS calculated with the optimal $\sigma$ = \unit[3.78]{bohr}. The local band gaps of Si and SiO$_2$ in the middle of the slab using the LMBJ ($E^{\mathrm{Si}}_g$ = \unit[1.39]{eV}, $E^{\mathrm{SiO}_2}_g$ = \unit[8.09]{eV}) compare well with the MBJ bulk values ($E^{\mathrm{Si}}_g$ = \unit[1.20]{eV}, $E^{\mathrm{SiO}_2}_g$ = \unit[8.79]{eV}, even we can clearly see that the SiO$_2$ layer is too thin to allow for the saturation of the local band gap to the correct bulk value. In fact, we can clearly observe in Fig.~\ref{fig:fig_heterostructure} that the interface states extend much more inside SiO$_2$ than in Si. In addition to the local band gaps, we can deduce the band offsets at the interface directly from our calculation. We obtain $\Delta E_{\mathrm{V}}$ = \unit[1.98]{eV} and $\Delta E_{\mathrm{C}}$ = \unit[4.72]{eV} for the valence and conduction band offsets, respectively. We should compare these numbers with experimental ($\Delta E^{\mathrm{exp}}_V$ = \unit[4.44]{eV} and $\Delta E^{\mathrm{exp}}_C$ = \unit[3.38]{eV}~\cite{Keister1999})
and theoretical $GW$ ($\Delta E^{GW}_V$ = \unit[4.1]{eV} and $\Delta E^{GW}_C$ = \unit[2.9]{eV}~\cite{Shaltaf2008}) and local hybrid 
values ($\Delta E^{LH}_V$ = \unit[4.27]{eV} and $\Delta E^{LH}_C$ = \unit[3.05]{eV}~\cite{Borlido2018}), the latter obtained for the same interface model as in our calculation. The comparison shows that LMBJ gives the correct type of band alignment, with both electrons and holes confined in the SiO$_2$ layer, however the bands of SiO$_2$ are placed $\sim$ \unit[2]{eV} too high in energy. This is a direct consequence of the fact that, for the Si/SiO$_2$ interface, the LMBJ potential gives basically the same valence band offset as PBE ($\Delta E^{\mathrm{PBE}}_V$ = \unit[2.13]{eV}). Finally, from the local band edges we obtain for the width of the interface $\sim$\unit[16.6]{bohr}, which is slightly larger than previous experimental~\cite{Muller1999} and theoretical~\cite{Yamasaki2001} results.

We complement the results for the Si/SiO$_2$ interface with band diagrams of other well studied semiconductor interfaces, presented in Tab.~\ref{tab:interfaces} in comparison with other theoretical and experimental results.
\begin{table*}[h!]
    \centering
    \begin{tabular}{|c | c|c|c|c|c|c|c|}
        \hline
          A/B heterostructure & & $E^{A}_g$ (het) & $E^{A}_g$ (bulk) & $E^{B}_g$ (het) & $E^{B}_g$ (bulk) & $\Delta E_v$ & $\Delta E_c$ \\
          \hline
          & PBE & 0.91 & 0.60 & 5.06 & 5.34 & 2.13 & 2.02\\
          & MBJ & 2.02 & 1.27 & 6.92 & 8.13 & 2.13 & 2.77\\          
        Si/SiO$_2$ & LMBJ & 1.39 & 1.20 & 8.09 & 8.71 & 1.98 & 4.72\\
          & $G_0 W_0$ &  & 1.23 &  &  & 4.1$^{a}$ & 2.9$^{a}$ \\
          & exp &  & 1.17 &  & 10.30$^{b}$ & 4.44$^{c}$ & 3.38$^{c}$\\
          \hline
          & PBE & 1.13 & 1.41 & 0.74 & 0.70 & 0.35 & 0.04\\
          & MBJ & 2.09 & 2.15 & 1.64 & 1.59 & 0.45 & 0.0\\           
        AlAs/GaAs & LMBJ & 2.03 & 2.13 & 1.80 & 1.59 & 0.63 & -0.40\\
          & $G_0 W_0$ &  & 2.09 &  & 1.32 & 0.60 & 0.17\\
          & exp &  & 2.23 &  & 1.52 & 0.53 & 0.18\\
          \hline
          & PBE & 1.49 & 1.56 & 1.53 & 1.61 & 0.28 & -0.31\\
          & MBJ & 2.48 & 2.37 & 2.25 & 2.38 & 0.48 & -0.25\\           
         AlP/GaP & LMBJ & 2.35 & 2.35 & 2.23 & 2.38 & 0.42 & -0.3\\
          & $G_0 W_0$ &  & 2.50 &  & 2.59 & 0.67 & -0.76\\
          & exp & &  2.51 & &  2.35 & 0.55 & -0.39\\
          \hline
          & PBE & 1.60 & 1.61 & 0.65 & 0.60 & 0.40 & 0.55\\
          & MBJ & 2.27 & 2.38 & 1.40 & 1.27 & 0.31 & 0.56\\           
         GaP/Si & LMBJ & 2.30 & 2.38 & 1.28 & 1.20 & 0.35 & 0.67\\
          & $G_0 W_0$ &  & 2.59 &  & 1.23 & 0.53 & 0.83\\
          & exp &  & 2.35 &  & 1.17 & 0.80 & 0.38\\
          \hline
    \end{tabular}
    \caption{Local band gaps $E_g$, valence ($\Delta E_v$) and conduction band ($\Delta E_c$) offsets in eV calculated with different XC potentials. All $G_0 W_0$ and experimental values are from Ref.~\citenum{Steiner2014}, unless stated otherwise. $^{a}$~Ref.~\citenum{Shaltaf2008}, $^{b}$~Ref.~\citenum{Marques2011}, $^{c}$~Ref.~\citenum{Keister1999}.
    }
    \label{tab:interfaces}
\end{table*}
For all the systems considered, using LMBJ we obtain local band gaps which agree well with bulk counterparts and experiment (MAPE = 0.11). As expected, PBE systematically underestimates local band gaps (MAPE = 0.41), while MBJ underestimates large band gaps and overestimates small band gaps in very heterogeneous systems (MAPE = 0.19), in particular Si/SiO$_2$. 

Concerning the band offsets, we encounter a different behavior of the LMBJ potential for different heterostructures in our limited test set. For Si/SiO$_2$ and GaP/Si $\Delta E^{\mathrm{LMBJ}}_V \approx \Delta E^{\mathrm{PBE}}_V \approx 0.5 \Delta E^{\mathrm{exp}}_V$. In this case, the error in $E^{\mathrm{LMBJ}}_{V}$ translates directly to $E^{\mathrm{LMBJ}}_{C}$ as the experimental band gap is correctly predicted. For the other two systems, AlP/GaP and GaP/Si, we obtain $E^{\mathrm{LMBJ}}_{V}$ close to the experimental value. For AlP/GaP LMBJ also yields a very good $E^{\mathrm{LMBJ}}_{C}$, while for GaP/Si it is slightly worse due to the overestimation of the band gap, leading even to a wrong interface type. A similar behavior for band offsets was obtained with hybrid functionals and $GW$ calculations~\cite{Alkauskas2011}, as well as with local hybrid functionals~\cite{Borlido2018}, although with an overall better performance than LMBJ for the four interfaces. We also find that the results are sensitive to the size of the supercell used to model the interface, as we are extracting band gaps from the value of the local band gap in the middle of the layer. Here we used interface models from the literature to enable comparisons with previous calculations. Those supercells were used originally in two-step calculations, where only the valence band offset was extracted from a supercell calculation, while band gaps were calculated for bulk crystals. We can conclude that one has to pay attention to include more atomic layers to extract accurate band diagrams from a single supercell calculation. In this respect, the use of LMBJ would become particularly advantageous when the supercell is large, due to its reduced computational cost in comparison with hybrid functionals or $GW$.

As a next step we consider the application of the LMBJ potential to a crystal with a surface. We chose as a test system the clean (001) surface of silicon with the $2\times 1$ reconstruction resulting from numerical optimization~\cite{Dabrowski1992}. This model agrees well with both experimental measurements~\cite{Over1997} and recent calculations~\cite{Seo2014}. We conducted the LMBJ calculation for a slab consisting of 32 atomic layers and the width of vacuum between periodic slabs was set to \unit[79.4]{bohr}. The energy cut-off was set to \unit[245.3]{eV} and we used an $8\times 8\times 1$ $\mathbf{k}$-point grid. As for the Si/SiO$_2$ interface, we chose for the smearing $\sigma=$ \unit[3.78]{bohr}, and a threshold density $\rho_{\mathrm{th}}$ corresponding to $r^{\mathrm{th}}_{s} = $ \unit[5]{bohr} as justified above.

\begin{figure} 
  \centering
  \includegraphics[width = 0.89\columnwidth]{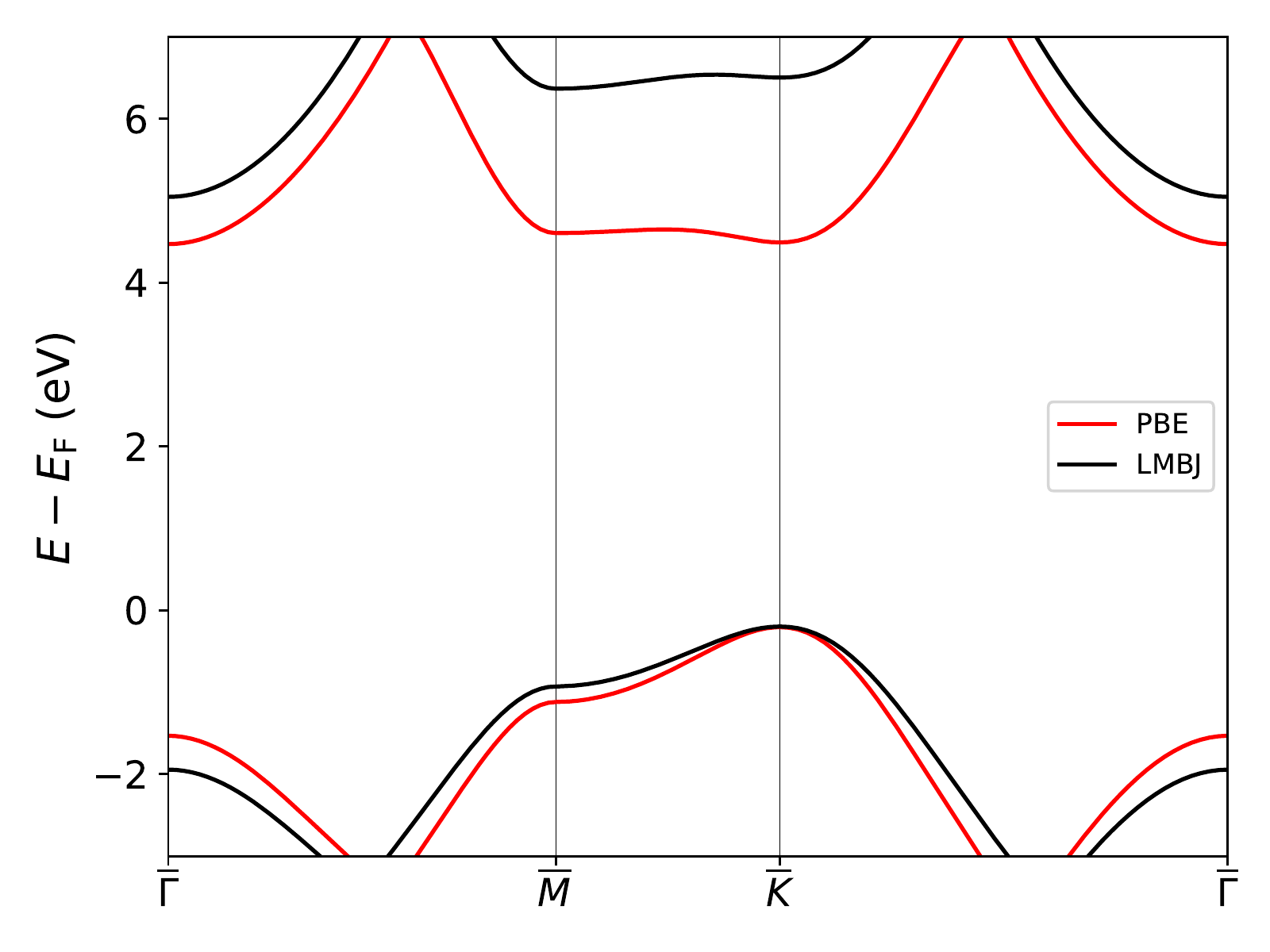}
  \includegraphics[width = 0.89\columnwidth]{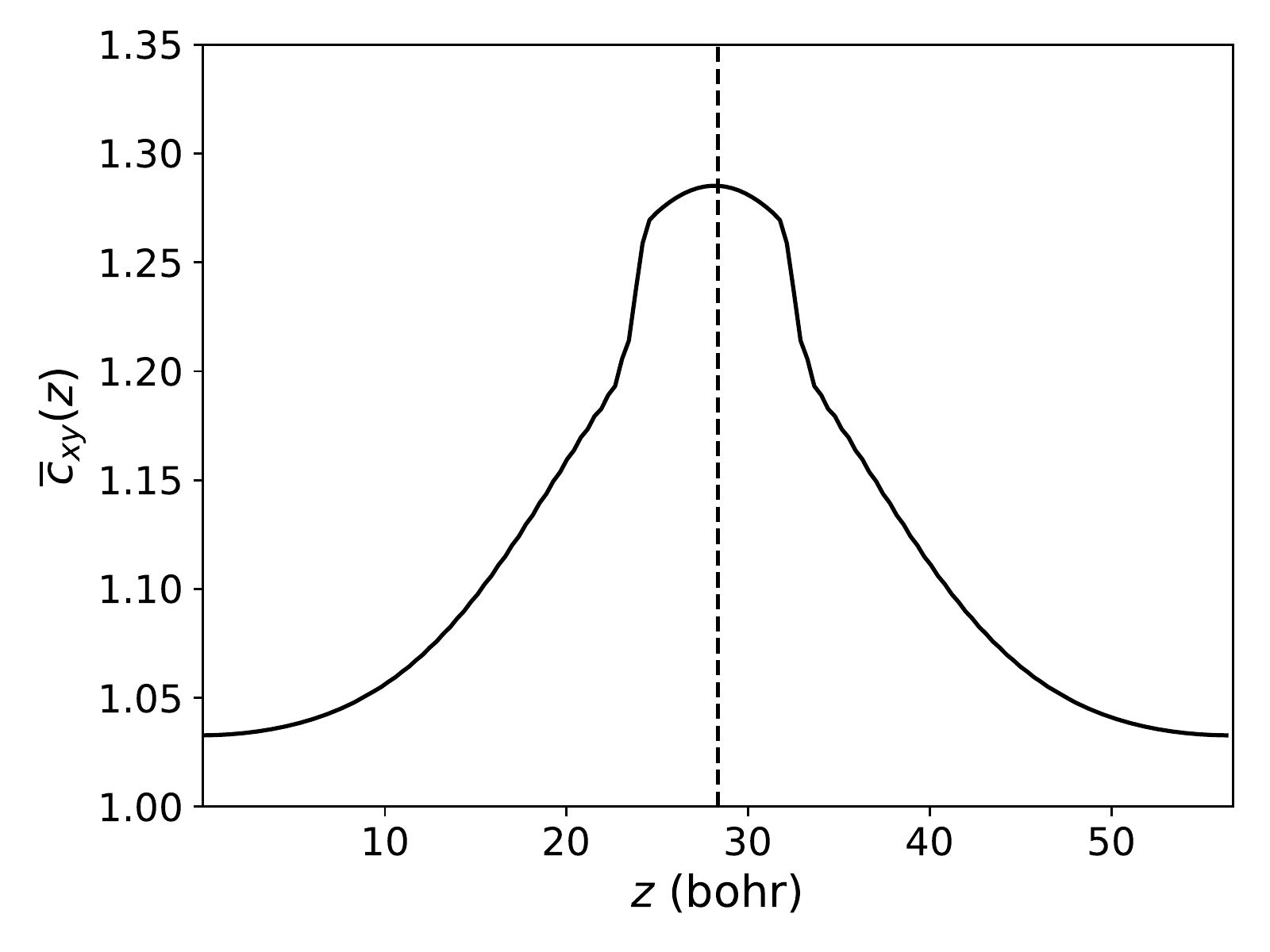}
  \caption{Hexagonal boron-nitride. Top: Band structure (highest valence and lowest conduction band) along high-symmetry lines in the Brillouin zone calculated with PBE and LMBJ. Bottom: Local mixing averaged in the $xy$-plane. The dashed vertical line denotes the position of the boron nitride atomic plane.}
  \label{fig:fig_BN}
\end{figure}

In Fig.~\ref{fig:fig_surf} (left) we show the converged average $\overline{c}_{xy}(z)$ along the direction perpendicular to the slab surface. We obtained $\overline{c}_{xy}(z)\approx 1.11$ for $z$ in the middle of the slab and $\overline{c}_{xy}(z)\approx 1.0$ for $z$ in the vacuum region. The former agrees well with the bulk mixing of Si ($c=1.13$) and the latter is the desired property of the LMBJ in vacuum. We observe a small peak exactly at the surface, which originates from the large gradient of the density and is thus of a physical origin. Even if varying $\rho_{\mathrm{th}}$ leads to a different size of this peak, these changes have no significant influence on the local potential, and thus on the electronic structure. This may be different in other materials and should be the subject of further investigations. In Figs.~\ref{fig:fig_surf} (middle) and (right) we present the LDOS and band structure of the Si slab. For the local band gap in the middle of the slab we obtained $E_g=$\unit[1.26]{eV}, which agrees well with the bulk calculation. Finally, the two surface states visible in both the LDOS and the band structure match with those calculated in Ref.~\citenum{Seo2014} using hybrid functionals, including their dispersion and distance from the bulk edges.

In our final test we turn to two-dimensional materials. We chose hexagonal boron-nitride (h-BN), since there is a renewed interest in its electronic structure~\cite{Geim2013,Withers2015} and the band gap of its parent bulk (three-dimensional) structure is described very well by the MBJ potential~\cite{Koller2012}. Bulk h-BN consists of layered honeycomb monolayers with in-plane lattice constant $a=$ \unit[4.72]{bohr} and interlayer distance of \unit[6.29]{bohr} (for the AA' stacking which was predicted to be most stable~\cite{Gu2007,Wickramaratne2018}).

We first calculated the electronic structure of bulk h-BN using the PBE, MBJ and LMBJ (with $\sigma=$~\unit[7.56]{bohr}) potentials with an energy cut-off of \unit[400]{eV} and a $21 \times 21 \times 17$ $\mathbf{k}$-point grid. We obtained band gaps $E^{\mathrm{PBE}}_g=$ \unit[3.88]{eV}, $E^{\mathrm{MBJ}}_g=$ \unit[5.64]{eV} and $E^{\mathrm{LMBJ}}_g=$ \unit[5.55]{eV}. The local mixing of the LMBJ calculation is basically constant in the bulk crystal: $c=1.31$ . This value is very close to the mixing $c=1.33$ of a bulk MBJ calculation. Both the MBJ and LMBJ band gaps agree well with the HSE and experimental results of $E^{\mathrm{HSE}}_g=$ \unit[5.95]{eV}~\cite{Wickramaratne2018} and $E^{\mathrm{exp}}_g=$ \unit[6.08]{eV}~\cite{Cassabois2016}, respectively. For the h-BN monolayer we kept the lattice constant $a=$ \unit[4.72]{bohr}. The width of vacuum between periodic replicas of the monolayers was set to \unit[56.7]{bohr}, the energy cut-off was \unit[400]{eV} and we used a $21 \times 21 \times 1$ $\mathbf{k}$-point grid. The smearing and threshold Wigner-Seitz radius were set to $\sigma=$~\unit[7.56]{bohr} and $r^{\mathrm{th}}_s=$~\unit[5]{bohr}, respectively. 

In the top part of Fig.~\ref{fig:fig_BN} we show the band structure calculated using the PBE and LMBJ potentials.
The indirect and direct (at $\overline{K}$) band gaps we obtained with PBE were $E^{\mathrm{PBE}}_g=$~\unit[4.68]{eV} and $E^{\mathrm{PBE}}_d=$ \unit[4.70]{eV}, respectively. This result is improved by the use of the LMBJ potential, which yields $E^{\mathrm{LMBJ}}_g=$~\unit[5.25]{eV} and $E^{\mathrm{LMBJ}}_d=$~\unit[6.70]{eV}. These values improve over PBE and differ by $\sim 10\%$ from values obtained by hybrid functional calculations ($E^{\mathrm{HSE}}_g=$~\unit[5.68]{eV}, $E^{\mathrm{HSE}}_d=$~\unit[6.13]{eV}~\cite{Haastrup2018}). Other theoretical and experimental works obtained a direct (indirect) band gap of \unit[6.47]{eV}~\cite{Wickramaratne2018} and \unit[6.1]{eV}~\cite{Elias2019}, respectively.

\section{Conclusions}
We proposed a generalization of the successful MBJ potential introduced by Tran and Blaha~\cite{Tran2009} to calculate bulk band structures. Our local MBJ potential enables the calculation of band diagrams of heterostructures and the evaluation of energy levels of finite systems. To this end, we defined a position dependent parameter $c\rdep$, averaged over a region of approximately one unit cell that replaces the constant parameter $c$ of the MBJ potential. We demonstrated that our LMBJ potential allows to obtain band diagrams at interfaces with other materials or with vacuum in a single calculation, reproducing well both surface states and bulk band states inside the layers. We discussed examples of the application of the LMBJ potential to semiconductor interfaces, a Si surface and a h-BN monolayer, demonstrating that it is possible to obtain band gaps of the quality of hybrid functionals even for a 2D material. A large-scale benchmark calculation is currently not possible due to the lack of reliable experimental data for lower-dimensional materials. Thanks to its computational efficiency, the LMBJ potential allows for reliable band structure calculations of large inhomogeneous systems, also when hybrid functional and $GW$ approaches are computationally too expensive.

\begin{acknowledgement}
  This work was supported by the Deutsche Forschungsgemeinschaft (DFG, German Research Foundation) through the projects SFB-762 (project A11), SFB-1375 (project A02), MA 6787/1-1 and BO 4280/8-1. S.B. and T.R. acknowledge funding from the Volkswagen Stiftung (Momentum) through the project ``dandelion''. 
\end{acknowledgement}

\begin{suppinfo}
The Supplemental Material includes details on the implementation of the LMBJ potential into the VASP code, the definition of the LDOS and results of band gap calculations for 3D bulk semiconductors.

\end{suppinfo}

\providecommand{\latin}[1]{#1}
\makeatletter
\providecommand{\doi}
  {\begingroup\let\do\@makeother\dospecials
  \catcode`\{=1 \catcode`\}=2 \doi@aux}
\providecommand{\doi@aux}[1]{\endgroup\texttt{#1}}
\makeatother
\providecommand*\mcitethebibliography{\thebibliography}
\csname @ifundefined\endcsname{endmcitethebibliography}
  {\let\endmcitethebibliography\endthebibliography}{}



\section*{Supplementary material}

# Local modified Becke-Johnson exchange-correlation potential for interfaces, surfaces, and two-dimensional materials (Supplemental Material)

Tomáš Rauch,[*,†] Miguel A. L. Marques,[‡,¶] and Silvana Botti[†,¶]

†*Institut für Festkörpertheorie und -optik, Friedrich-Schiller-Universität Jena, Max-Wien-Platz 1, 07743 Jena, Germany*

‡*Institut für Physik, Martin-Luther-Universität Halle-Wittenberg, 06120 Halle/Saale, Germany*

¶*European Theoretical Spectroscopy Facility*

E-mail: tomas.rauch@uni-jena.de

# 1 Implementation of the local modified Becke-Johnson potential into the VASP code

We implemented the LMBJ XC potential into the VASP code,[1] which uses the projector-augmented-waves (PAW) method[2] to describe the interaction of the electrons with ions. For the original MBJ potential the average estimator $\bar{g}$, Eq. (4) of the main text, is evaluated inside the augmentation spheres on radial grids in addition to the plane-wave part. This is necessary, since the total contribution from the augmentation spheres amounts ∼10 % of the total $\bar{g}$. The same value of $\bar{g}$ is then used for the XC potential on all grid points of both the

1

real-space and radial grids for the plane-wave and augmentation parts, respectively.

In our local version of the MBJ potential, on the other hand, both the plane-wave and augmentation parts have to maintain their local character. Therefore, our calculation consists of the following steps:

1. $g_{\text{PW}}(\boldsymbol{r})$ for each $\boldsymbol{r}$ on the real-space grid and $g_{\text{PAW}}^i$ for each ion is calculated. Note that $g_{\text{PAW}}^i$ is already the mean value of $g$ corresponding to the $i$-th augmentation sphere. We chose this simplification, since we expect $g$ to be almost constant in near-atom regions after averaging (see step 2.).

2. Local averaged estimators are calculated. For the plane-wave part the integral over the unit cell
$$\bar{g}_{\text{PW}}(\boldsymbol{r}) = \int d^3r' G(\boldsymbol{r} - \boldsymbol{r}') g_{\text{PW}}(\boldsymbol{r}') \tag{1}$$
is evaluated using a Fourier transform and for the augmentation part a direct weighted sum over all ions,
$$\bar{g}_{\text{PAW}}^i = \sum_j G(\boldsymbol{r}_i - \boldsymbol{r}_j) g_{\text{PAW}}^j, \tag{2}$$
is calculated, where $G(\boldsymbol{r})$ is a Gaussian, see Eq. (5) of the main text.

3. The augmentation part is added to the plane-wave one and vice versa. This is done again by summing over all grid points weighted by the Gaussian:
$$\bar{g}_{\text{PW}}(\boldsymbol{r}) \rightarrow \bar{g}_{\text{PW}}(\boldsymbol{r}) + \sum_i G(\boldsymbol{r} - \boldsymbol{r}_i) g_{\text{PAW}}^i \tag{3}$$
$$\bar{g}_{\text{PAW}}^i \rightarrow \bar{g}_{\text{PAW}}^i + \int d^3r G(\boldsymbol{r} - \boldsymbol{r}_i) g_{\text{PW}}(\boldsymbol{r}). \tag{4}$$

Following these steps the LMBJ potential can be evaluated for a system without vacuum. If there is vacuum in the unit cell, like in surface calculations, two additional modifications are necessary. First, $g_{\text{PW}}(\boldsymbol{r})$ in step 1 is calculated using Eq. (7) in the main text. Second, the volume corresponding to each augmentation sphere $V_{\text{PAW}}^i$ has to be known to calculate

the mean $g^i_{\text{PAW}}$ for each augmentation sphere. Following the implementation of the original MBJ potential, this is approximated by $V^i_{\text{PAW}} = V_{\text{PAW}} = V_{\text{cell}}/N$, where $N$ is the number of ions in the system. This approximation has to be altered to $V_{\text{PAW}} = (1 - r_{\text{v}})V_{\text{cell}}/N$ for systems with vacuum, where $r_{\text{v}}$ is the ratio of the volume occupied by vacuum to the full unit cell volume. The ratio $r_{\text{v}}$ is obtained during each self-consistency cycle by evaluating the local charge density $\rho(\boldsymbol{r})$ at each grid point and comparing it with a threshold density $\rho^{\text{th}}$ corresponding to the Wigner-Seitz radius $r_s^{\text{th}}$ defined in the main text. The quantity $r_{\text{v}}$ is then given as the ratio of the number of grid points with $\rho(\boldsymbol{r}) < \rho^{\text{th}}$ to the number of grid points with $\rho(\boldsymbol{r}) > \rho^{\text{th}}$.

## 2 Local density of states (LDOS)

To evaluate the local electronic structure, we calculate the local density of states (LDOS) $D_i(\epsilon)$, where we divided the unit cell into slices $\Omega_i$ parallel to the interface with thickness $\Delta z$. The LDOS is then obtained as[3,4]

$$D_i(\epsilon) = \frac{1}{V_{\text{BZ}}} \sum_n \int_{\text{BZ}} \omega^i_{\mathbf{k}n} \delta(\epsilon - \epsilon_{\mathbf{k}n}) d^3\mathbf{k} \tag{5}$$

with the weight function

$$\omega^i_{\mathbf{k}n} = \int_{\Omega_i} |\varphi_{\mathbf{k}n}(\boldsymbol{r})|^2 d^3\mathbf{r}, \tag{6}$$

where $\varphi_{\mathbf{k}n}(\boldsymbol{r})$ are the Kohn-Sham wave functions at wave vector $\mathbf{k}$ of band $n$ with eigenvalue $\epsilon_{\mathbf{k}n}$.

## 3 Band gaps of bulk 3D semiconductors

In Fig. 1 we show band gaps calculated with the LMBJ potential for different values of $\sigma$. While the band gap varies strongly for smaller $\sigma$, it saturates at $\sigma \approx 3.78\,\text{bohr} = 2.0\,\text{Å}$, where the calculated band gap values of the original MBJ potential are restored again.

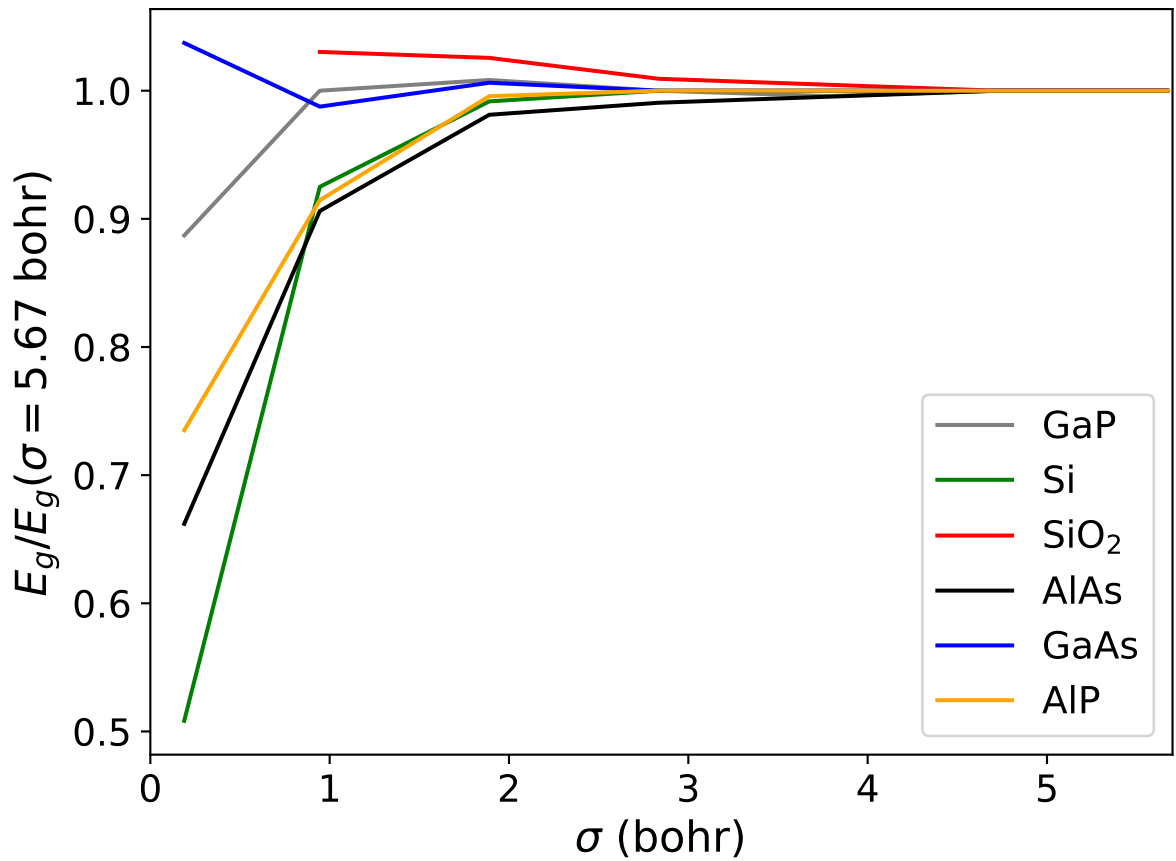

Figure 1: Band gaps of chosen semiconductors calculated with the LMBJ potential with $r_s^{\text{th}} = 5.0$ bohr and varying $\sigma$.



\end{document}